# CHARACTERIZING AND EVALUATING THE IMPACT OF SOFTWARE INTERFACE CLONES

Hani Abdeen[1] and Osama Shata[2]

[1,2]Computer Science Engineering Department, Engineering College – Qatar University,
P.O.Box 2713, Doha, Qatar
`hani.abdeen@qu.edu.qa` -- `sosama@qu.edu.qa`

## ABSTRACT

*Software Interfaces are meant to describe contracts governing interactions between logic modules. Interfaces, if well designed, significantly reduce software complexity and ease maintainability. However, as software evolves, the organization and the quality of software interfaces gradually deteriorate. As a consequence, this often leads to increased development cost, lower code quality and reduced reusability. Code clones are one of the most known bad smells in source code. This design defect may occur in interfaces by duplicating method/API declarations in several interfaces. Such interfaces are similar from the point of view of public services/APIs they specify, thus they indicate a bad organization of application services. In this paper, we characterize the interface clone design defect and illustrate it via examples taken from real-world open source software applications. We conduct an empirical study covering nine real-world open source software applications to quantify the presence of interface clones and evaluate their impact on interface design quality. The results of the empirical study show that interface clones are widely present in software interfaces. They also show that the presence of interface clones may cause a degradation of interface cohesion and indicate a considerable presence of code clones at implementations level.*

## 1. INTRODUCTION

Software Interfaces represent abstract service contracts governing interactions between logic modules. Formally, they are reference types used to encode similarities among classes of different types. They are also used to define the system Application Programming Interfaces (API). Well-designed interfaces significantly ease program comprehension, reduce software complexity and ease maintainability by fostering modularization, hiding implementation details and minimizing the impact caused by changes in the software implementation [10]. A good design of interfaces is a fundamental key to understand the whole application services, and the interactions between modules and subsystems.

However, as software evolves over time with the modification, addition and removal of new classes, interfaces and services/APIs, the software gradually drifts and looses quality [7]. As a consequence, the organization and the quality of the implementation, and so of interfaces, gradually deteriorate [7]. The presence of interface design defects often leads to increased development cost, lower code quality, reduced reusability and reduced development productivity [23, 24]. Code clones are one of the most known bad smells in source code [8, 1]. Although interfaces do not provide implementations, code clones may still occur by duplicating method/API declarations in several interfaces. Such interfaces that share identical method declarations are thus similar from the point of view of public services/APIs they specify. Hence, they indicate a bad organization of application services.

Despite the importance of interfaces, currently there are only a few research efforts that investigate the quality of interfaces design and systematically assess their adherence to well-known interface design principles. Most existing work largely investigated the detection of design anomalies at the class level without focusing on the specifics of interfaces [21]. Studying the impact of interface design anomalies on software quality has been mostly neglected. This paper aims to address this gap as it constitute a potential barrier towards fully realizing the benefits of using interfaces as a central element for modular design.

**Contributions**. In this paper, we investigate and characterize the design defect related to interface clones (*i.e.*, interfaces with duplicate methods where some interface methods are declared several times in different interfaces). We illustrate this design anomaly via examples taken from real-world open source applications. We present an empirical study covering 9 open source projects to evaluate the quality of interfaces using qualitative and quantitative analysis of the source code in order to quantify the presence of interface clones and estimate their impact on the interface design quality. The results of our study show that interface clones are widely present in software interfaces. They also show that the presence of interface clones causes the degradation of interface usage cohesion and indicates a considerable presence of code clones at implementations level.

The remainder of the paper is organized as follows. Section 2 discusses existing works related to software interface design quality. We characterize the interface clones design defect via examples taken from real-world software applications in Section 3. Then in Section 4 we describe the empirical study conducted on 9 real world open source software applications to study the interface clones design defect and answer different research questions. Section 5 presents and analyses the results of the empirical study before concluding.

## 2. RELATED WORKS

In recent years there has been considerable interest in automatic detection of design defects in object oriented software [8, 18, 22, 13, 19]. Mens and Tourwé [15] survey shows that existing approaches are mainly based on code metrics and predefined bad smells in source code [5, 9, 2, 8]. Unfortunately, none of those code smells and relevant OO metrics, such as metrics of Chidamber and Kemerer (CK) [5], are applicable to software interfaces [21] –since interfaces do not contain any logic, such as method implementations, invocations, or attributes.

Few recent works attempt to address the particularities of interfaces. Boxall and Araban define a set of primitive metrics to measure the complexity and usage of interfaces [3]. The authors in [26] define a new list of metrics for assessing the design quality of interfaces with regard to the redundancy in interface hierarchies and to the similarities between interfaces. Martin proposed the Interface Segregation Principle (ISP) [14]. Hence, an interface should group methods that are used together to serve a specific client. Large interfaces should be split into smaller more specific ones so that clients will only have to know relevant methods without unwanted coupling to those that they do not use [14, 21]. Ideally, an interface should not expose any unused methods and all the declared methods should be used by every client of the interface.

With regard to the ISP principle, Romano and Pinzger [21] used the Service Interface Usage Cohesion (SIUC) to measure the violation of ISP. SIUC is defined by Perepletchikov [20], and referred to by Romano and Pinzger [21] as Interface Usage Cohesion (IUC): $IUC(i)$ computes the usage cohesion of *i* with regard to all its client classes. IUC states that an interface has a strong cohesion if every client class of that interface uses all the methods declared in it. IUC takes its value in the interval [0..1]. The larger the value of IUC, the better is the interface usage cohesion.

Besides Martin's design principles ISP and DIP [14], Boxall and Araban's primitive metrics [3], Abdeen and Shata's interface design metrics [26], and Perepletchikov's SIUC metric [20], there are no tools that help in evaluating the impact of interface clones.

The authors in [21] investigated the suitability of existing source code metrics to classify Java interfaces into change-prone and not change-prone. The metrics used in the study were Chidamber and Kemerer (CK) metrics [5], interface complexity metrics [3] and the IUC metric [20]. They empirically evaluated their model for predicting change-prone Java interfaces by investigating the correlation between metrics and the number of changes in interfaces of ten Java open-source systems. The paper concluded that most of the CK metrics are not sound for interfaces and only perform well for predicting change-prone concrete and abstract classes. Therefore this confirms the claim that interfaces need to be treated separately. The IUC metric exhibits the strongest correlation with the number of interface changed. Hence IUC can improve the performance of prediction models for classifying Java interfaces into change-prone and not change-prone.

Kawrykow and Robillard [11] report a technique and tool support to automatically detect patterns of poor API usage in software projects by identifying client code needlessly re-implementing the behavior of an API method. Their approach uses code similarity detection techniques to detect cases of API method imitations then suggest fixes to remove code duplication. The paper addressed a specific scenario of duplicate API implementation by focusing on the analysis of the client code. Our approach is different and relies on the analysis the interfaces themselves to detect clones between software interfaces and evaluate their impact of interface design quality.

The work in [4] examined the relative impact of interface complexity (e.g. interface size and operation argument complexity) on the failure proneness of the implementation using data from two large-scale systems. This work provides empirical evidence that the increased complexity of interfaces is associated with the increased failure proneness of the implementation (e.g., likelihood of source code files being associated with a defect) and higher maintenance time. Our works goes much further by studying specific interface design defects and investigating their impact on interface quality.

## 3. INTERFACE CLONES

In this section we characterize and describe the interface clone design anomaly via real examples taken from real-world Java applications. Before detailing the design anomaly, we introduce the vocabularies we use in the rest of this paper.

Interface: we consider an interface as a set of method declarations (*i.e.*, a set of signatures). In this paper we do not take into account marker interfaces or interfaces declaring only constants. We consider only signatures that are explicitly declared as "public". We define the size of an interface *i*, *size(i)*, by the number of public methods that *i* declares.

Identical Signatures/Methods within Interfaces: we say that two method signatures are identical if they have the same return-type, the same name and the same list-of-parameter-types.

When identical signatures are present between different interfaces then they represent duplicate interface methods and cause interface clones.

### Definitions:

- *Interface Duplicate Method* represents a public method signature that is declared (duplicated) within several interfaces. The set of methods that are declared in an interface *i* and are duplicated in other interfaces are referred by Interface duplicate methods in *i*, IDM(*i*).
- We say that there is an *Interface Clone* between two interfaces, *i* and *j*, if there is a shared duplicate methods between them –*i.e.*, *i* and *j* both declare an identical subset of methods.
- A *Duplicate Interface* is an interface that all its methods are re-declared (cloned) in one or more interface(s).

## 3.1. Interface Duplicate Methods

Fig. 1 shows an example of interface clones, taken from *Vuze* software system. It shows a group of three interface methods that are declared several times in different interfaces and cause interface clones between five interfaces: (A) *DiskManagerWriteRequest*, (B) *DiskManagerReadRequest*, (C) *PeerReadRequest*, (D) *DiskManagerReadRequest* and (X) *DiskManagerWriteRequest* in *core3::disk* namespace. Moreover, the figure shows that the interfaces A and B represent duplicate interfaces. Furthermore, interface clones are included in three other interfaces C (*PeerReadRequest*), D (*DiskManagerReadRequest*) and X (*DiskManagerWriteRequest* in *core3::disk* namespace). Note that D extends C.

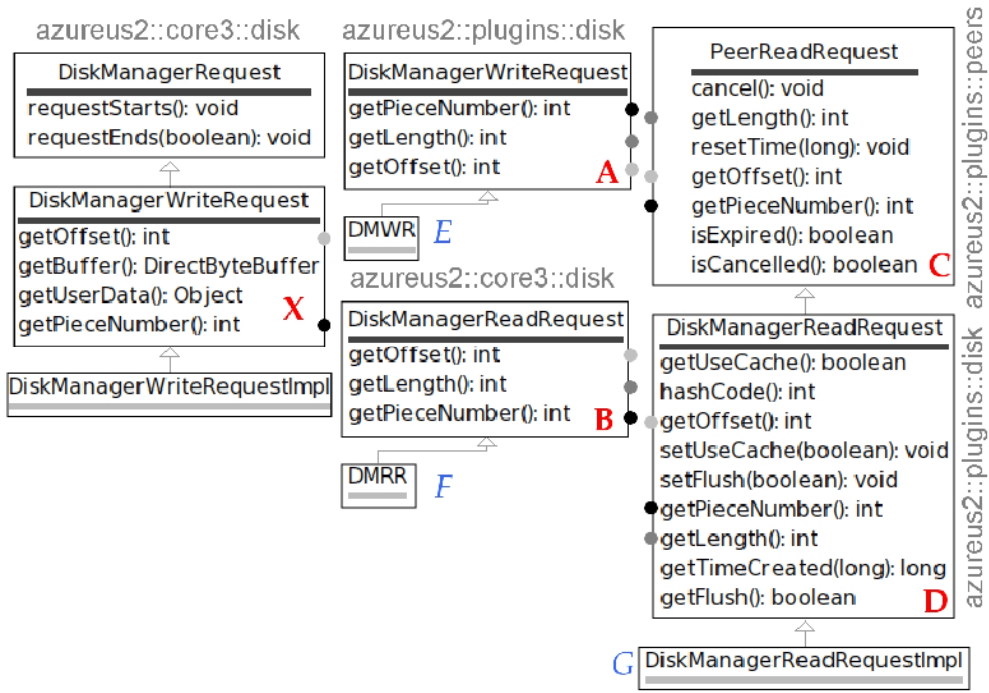

Figure 1: Example of Interface Similarity.

Duplication of read & write request methods in 5 interfaces in *Vuze* software system.

A conclusion is that the read and write request methods in *Vuze* application are declared in four interfaces, and two of them are declared in five interfaces. These duplicate declarations add more complexity and needlessly increase the size to the software system. For example, to locate the read and write request methods in *Vuze*, one needs to locate five interfaces and inspect the sub-hierarchy of each of those interfaces. Instead, those methods should be declared only once. In the following, we propose a possible refactoring for reducing the declaration redundancy of read and write methods in *Vuze*. First of all, we can safely remove interface A (*DiskManagerWriteRequest*) after moving all its incoming dependencies to interface B (*DiskManagerReadRequest*), e.g., make the class *DMWR* (E) implement interface B instead of A. It is worth noting that classes *DMWR* (E) and *DMRR* (F) both provide identical implementations for those duplicate methods. We then propose to rename B to DiskManagerRWRequest –since it declares what is apparently R&W methods. Then, we propose the following refactoring: 1) rename interface D (*DiskManagerReadRequest*) to ...*PeerReadRequest*; 2) make C (*PeerReadRequest*) a sub-interface of B (*DiskManagerReadRequest*); 3) remove the set of those duplicate signatures from both C (*PeerReadRequest*) and D (*DiskManagerReadRequest*). The result of our proposed refactoring is that the R&W methods are declared only in interface B (*DiskManagerReadRequest*). Hence the number of interfaces is reduced –A is removed; E and F, which need to implement the same set of methods, both implement B; and the size of interfaces C and D is reduced.

## 4. STUDY SETUP

This section describes the setup of the empirical study that we conduct in this paper around the interface clone design anomaly.

### 4.1. Research Questions

The research questions that the empirical study aims at answering are listed as follows:

**Q1** *To what extent interface clones are present in real software applications?*
**Q2** *Do interface clones help in improving the interface usage cohesion?*
**Q3** *Are interface clones a symptom of fat interfaces?*
**Q4** *To what extent code clones are present between the implementations of cloned interface methods?*

### 4.2. Case Studies

Table 1: Information about case-study software applications.

|  | **\|C\|** | **LC** | **\|I\|** | **sum *size(i)*** |
|---|---:|---:|---:|---:|
| *jFreeChart* | 590 | 160k | 94 | 560 |
| *JHotDraw* | 667 | 65k | 46 | 537 |
| *GanttProject* | 886 | 44k | 106 | 587 |
| *Plantuml* | 1224 | 76k | 122 | 358 |
| *ArgoUML* | 2202 | 166k | 126 | 1442 |
| *Opentaps* | 3028 | 416k | 166 | 1598 |
| *Rapidminer* | 3660 | 222k | 142 | 633 |
| *Hibernate* | 6195 | 373k | 471 | 3185 |
| *Vuze* | 6564 | 635k | 933 | 6476 |

|C|   number of classes (not interfaces); |I|   number of interfaces; sum *size(i)*   number of all declared public methods in all interfaces.

To study the interface clone design anomaly and answer the above questions, we conducted an empirical study on nine Java applications (Table 1): 1- *jFreeChart$_{v1.0.14}$* , 2- *JHotDraw$_{v7.1}$* , 3- *GanttProject$_{v2.1}$* , 4- *ArgoUML$_{v0.28.1}$* , 5- *Plantuml$_{v7935}$* , 6- *Rapidminer$_{v5.0}$* , 7- *Opentaps$_{v1.5.0}$* , 8- *Hibernate$_{v4.1.4}$* , 9- *Vuze$_{v4700}$* . We used the Moose tools-suit for data and software analysis [6] to parse the application source-code and identify interface clones. Note that the information in Table 1 about the studied applications are obtained after excluding the following: library interfaces; marker interfaces (*i.e.*, |*size(i)*| = 0); and test interfaces (*i.e.*, interfaces that are packaged in 'test(s)' packages and/or all their implementations are test classes).

## 5. RESULTS

This section presents and analyses the results of the empirical study. Each subsection addresses one of the research questions that are outlined in Section 4.1.

### 5.1. Presence of Interface Clones

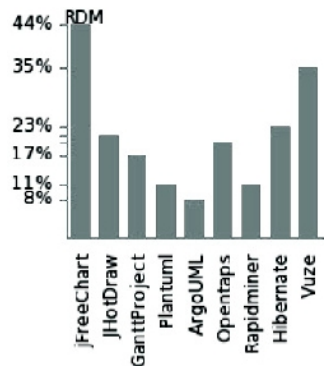

Figure 2: The Ratio of Interface Duplicate methods (RDM) within studied applications.

To answer Q1, we use DM to denote the number of all interface methods that are duplicate within the studied application (DM = $\sum$ IDM (*i*) $\forall i \in$ I). Where IDM(*i*) returns the number of duplicate methods in *i*. Using the IDM measurement we define the Ratio of Duplicate Methods: RDM = DM/$\sum$ *size(i)* $\forall i$. RDM takes its value in [0..1]. The largest is the value of RDM, the largest is the set of duplicate methods in the software interfaces of the system under analysis, thus the largest is the set of interface clones.

Fig. 2 shows that interface duplicate methods are present, surprisingly to considerable extents, in all 9 studied applications. This statistically shows the uncomfortable evidence of the presence of the interface clone design defect that this paper presents. It shows that 8% of *ArgoUML* interface methods and 44% of *jFreeChart* interface methods are redundantly declared (duplicated) within the application interfaces, and thus represent interface clones.

### 5.2. The Relation Between Interface Cohesion and Interface Clones

Fig. 3 shows that there is always a negative correlation between the number interface duplicate methods IDM(*i*) and the interface usage cohesion IUC(*i*). It shows that for 7 case-studies, against 2 ones, there is at least a moderate negative correlation (*i.e.*, the correlation is smaller than −0.35) between IDM(*i*) and IUC(*i*). This leads to the empirical conclusion that interface duplicate methods (*i.e.*, interface clones) have a negative effect on the interface cohesion. Thus, to answer Q2, the empirical results show that if the set of duplicate methods within an interface *i* gets larger, then the *i*' usage cohesion deteriorates and gets worst. As a consequence, we can

International Journal of Software Engineering & Applications (IJSEA), Vol.4, No.1, January 2013

state that duplicating methods in interfaces does not help in improving the interface usage cohesion. As a summary: interfaces that are characterized by high values of IDM(*i*) are probably characterized by a small values of IUC(*i*). Reducing interface clones will improve the quality of interface usage cohesion.

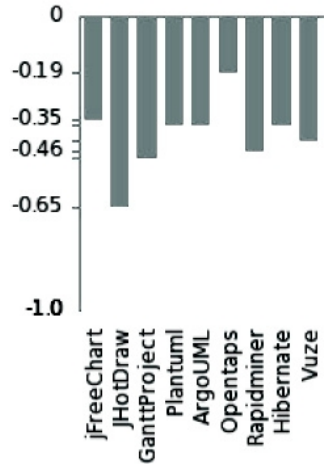

Figure 3: The correlation between Interface Cohesion IUC(*i*) and the number of interface duplicate methods IDM(*i*).

## 5.3. The Relation Between Interface Size and Interface Clones

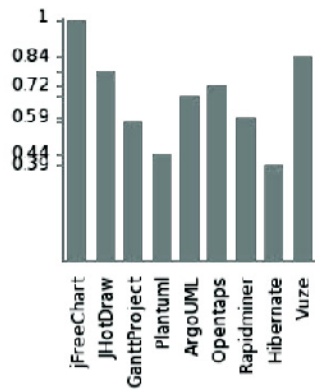

Figure 4: The correlation between the Interface size *size(i)* and the number of interface duplicate methods IDM(*i*).

Fig. 4 shows that there is always a strong-to-high correlation between the interface size and the number of interface duplicate methods: for 7 case-studies, against only 2, the correlation coefficient is larger than 0.58. This confirms the expected conclusion that the larger is the interface size, the larger is the probability to have more and more duplicate methods in it (*i.e.*, interface clones related to it). To answer Q3, on one hand Fig. 4 shows that interface clones are mostly a symptom of large interfaces. On another hand, by computing the determination of the correlation coefficient, which is simply the square of the correlation coefficient, we find that apart from the interface size, there are other factors which also influence the number of interface duplicate methods. Fig. 4 shows that the correlation coefficient between *size(i)* and IDM(*i*) for most of the case-studies is larger than 0.58. Hence, the coefficient determination is larger than 0.34 ($0.58^2$ ). This means that the variability in *size(i)* account for only 34% of the variability in IDM(*i*). As a consequence, we can say that interface duplicate methods could be found in small interfaces as in larger ones, even though their number in large interfaces is probably larger than in smaller interfaces. These findings imply that software developers have to think twice before adding new declarations to their interfaces. Interfaces should be designed as small as possible, not only to meet the ISP [14, 21], but also to avoid method duplication by reusing (*i.e.*, extending) existing software interfaces.

### 5.4. The Relation Between Interface Clones and Code Clones

This section aims at answering Q4. For this purpose, we used the *SmallDude*[1] tool for scanning the code of case-study applications and detecting code clones among the implementations of interfaces' duplicate methods. Fig. 5 shows our approach for associating between code clones and interface clones. In Fig. 5 the detected code clones between the implementations groups {*ClassA1.foo()*, *ClassA2.foo()*} and {*ClassB1.foo()*}, are associated to the interfaces *InterfaceA* and *InterfaceB*. Similarly for the detected code clones between the implementations groups {*ClassA1.do()*, *ClassA2.do()*} and {*ClassB1.do()*}.

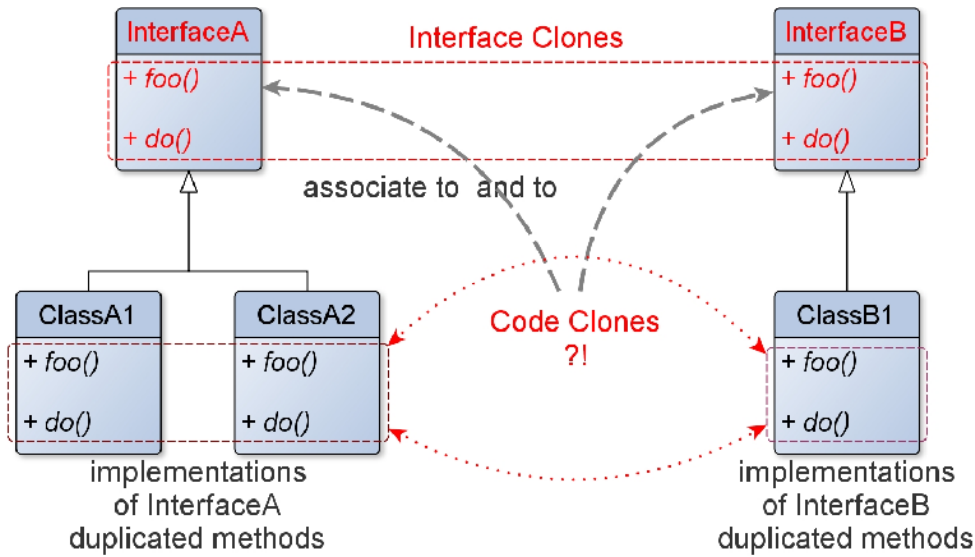

Figure 5: Explanation of our approach to associate code clones to interface clones.

Answering Q4 is not a trivial task. On one hand, Fig. 6 shows that code clones do exist always between the implementations of duplicate interface methods. It shows that there is always a considerable number of duplicated lines of code that are associated to Interface Clones. For example, in *jFreeChart*, *Opentaps* and *Hibernate*, there are more than 1k of LC that are involved in the code clones between the implementations of interfaces' duplicate methods. On another hand, Fig. 7 shows that the correlation between the number of interface duplicate methods DM(*i*) and the number of interface associated code clones is not conclusive. In some case-studies, such as *JFreeChart* and *Opentaps*, the correlation is strong, while it is poor in almost all other case-studies. This means, statistically speaking, that there is no decisive correlation between the number of interface duplicate methods and the number of code clones within the implementations of interfaces' duplicate methods. However, Fig. 7 shows that there is always a positive correlation between those two variables. This indecisive finding can be somewhat mitigated by the following facts:

- Parameters names can differ in the implementations of an interface method. An interface method declaration declares the parameter types and delegates the responsibility of specifying the parameter names to its implementations. Thus, the implementations of duplicate interface methods can specify different parameter names. This fact mitigates the correlation between code clones and method clones in interfaces.

- Unlike method declarations in interfaces, which are either identical or not, the number and/or the size of code clones within method implementations depend on the size (*i.e.*, number of lines of code) of those implementations. This implies that two large implementations can effectively involve larger number of code clones than many small implementations.

- Furthermore, an interface method can have one or many implementations. As a consequence, the relation between the number of interface cloned methods and the number of associated code clones, if any, can likely be non-linear.

---

[1] SmallDude (http://www.moosetechnology.org/tools/smalldude) is a text-based, language-independent code duplication (code clones) detector integrated within the Moose platform. The SmallDude' algorithm is a string based one that depends on 3 parameters: (1) Minimum clone length - defines the minimum amount of lines present in a clone. In our empirical study this parameter is set to 6. (2) Maximum line bias - defines the maximum amount of lines in between two exact chunks. In our empirical study this parameter is set to 2. (3) Minimum chunk size - defines the minimum amount of lines of an exact chunk. In our empirical study this parameter is set to 3.

As a summary, this section answers partially Q4 by demonstrating that code clones are always present between the implementations of cloned methods in interfaces as shown in Fig. 6. However, it does not conclude that code clones are more likely to occur within those implementations than within the implementations of interface methods that are not duplicated. We believe that further investigations of interface method clones and code clones are required to elaborate a conclusive response to Q4.

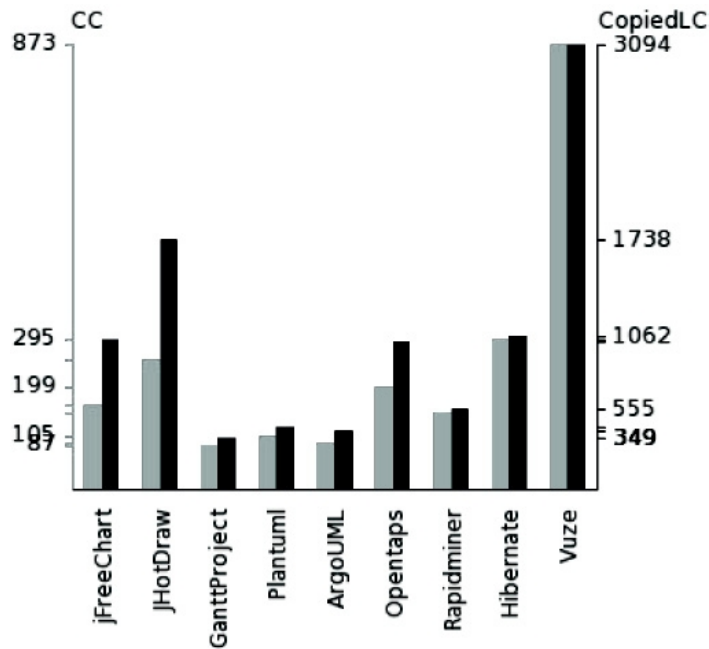

Figure 6: The number of code clones (CC) and sum of copied lines of code (CopiedLC) that are associated to Interface duplicate methods.

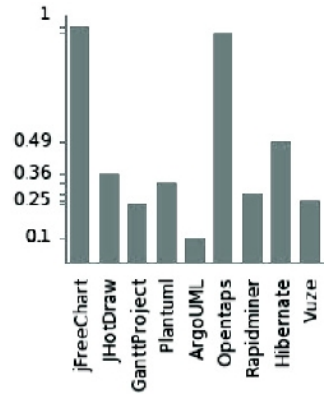

Figure 7: The correlation between the number of duplicate interface methods and the number of associated code clones.

## 6. CONCLUSION AND FUTURE WORK

In this paper, we characterized the interface clones design defect and illustrated it via real examples taken from real world software applications. We evaluated the impact of interface clones on interface design quality through an empirical study covering nine open source projects. The empirical results showed that interface clones are reliable symptoms of poor interface design. The results also show that interface clones are present, to different degrees, in interfaces of real systems. Our findings suggest that software development tools should be enhanced to detect interface clones and suggest appropriate refactoring of interfaces. This can assist software engineers to avoid interface clones so as to improve the design quality of interfaces and avoid code clones within the implementations of interfaces. A direction of future work is to define new quality metrics that assist in detecting different interface design anomalies and extract candidate interfaces for refactoring.

### ACKNOWLEDGMENT

This publication was made possible by NPRP grant #09-1205-2-470 from the Qatar National Research Fund (a member of Qatar Foundation). The statements made herein are solely the responsibility of the author.

International Journal of Software Engineering & Applications (IJSEA), Vol.4, No.1, January 2013